\documentclass[12pt,a4paper]{article}

\usepackage{amsmath, amssymb}

\begin{document}

\title{Comment on \lq\lq Doppler signature in electrodynamic retarded potentials,\rq\rq\ 
by Giovanni Perosa, Simone Di Mitri, William~A.~Barletta, Fulvio Parmigiani 
[{\it Physics Open} {\bf 14}, 100136 (2023)] }
\author{C\u alin Galeriu \\ Mark Twain International School, Voluntari, IF, Romania}

\maketitle

The authors of Ref. \cite{Perosa2023} have declared that they have offered 
\lq\lq a direct physical interpretation of the term ($v/c$)\rq\rq\  
characterizing the electromagnetic retarded potentials, and that their derivation
\lq\lq makes evident the physical phenomenology from which the electrodynamics
L-W potentials originate\rq\rq.
These claims deserve further clarifications.

The  Doppler factor in the Li\'{e}nard-Wiechert (L-W) potentials appears as the result of using a 
Dirac delta function identity 
\begin{equation}
\delta\big( g(x) \big) = \sum_n \frac{\delta(x - x_n)}{|g'(x_n)|},
\label{eq:Dirac_delta_property}
\end{equation}
where $g(x_n) = 0$ and the derivatives $g'(x_n) \neq 0$,
during the calculation of the electromagnetic potentials.

This derivation is most clearly presented in the book by Zangwill \cite{Zangwill2013}.
For a point particle with electric charge $q$, position ${\bf r}_0(t')$ and velocity ${\bf v}(t')$ at time $t'$,
the retarded L-W electromagnetic potentials are calculated as
\begin{equation}
\Big( \Psi({\bf r}, t), {\bf A}({\bf r}, t) \Big)
= q \int \frac{\delta(t' - t + R(t')/c)}{R(t')} \, \Big( 1, \frac{{\bf v}(t')}{c} \Big) \, dt',
\label{eq:Green_fn_solution}
\end{equation}
where $R(t') = |{\bf r} - {\bf r}_0(t')|$.
In our case the function $g(t')$ is 
\begin{equation}
g(t') = t' - t + \frac{R(t')}{c},
\label{eq:function}
\end{equation}
and the retarded time $t_r$ is the unique solution for which $g(t_r) = 0$.
We also notice that $g'(t_r) > 0$.

Instead of using the well known identity (\ref{eq:Dirac_delta_property}), 
the authors derive it, in a convoluted manner, with the help of a Fourier representation
of the Dirac delta function. 
The actual derivation of the relevant Dirac delta function identity is embedded
deep inside the standard calculation of the Li\'{e}nard-Wiechert potentials, 
the two derivations are thoroughly entangled together, and to the unsuspecting eye it is not
very easy to see what is going on. 

Consider a given test function $f(t)$, and the integral
\begin{equation}
\int \delta\big( g(t) \big) \ f(t) \ dt.
\label{eq:integral}
\end{equation}
How could we calculate this integral without a direct use of the identity (\ref{eq:Dirac_delta_property})?
A standalone derivation of this identity, 
in the simple case when the function $g(t)$ has only one root $t_r$, 
with $g(t_r) = 0$ and $g'(t_r) \neq 0$, consists
of two steps.

Step I. For the given test function $f(t)$, the domain of integration of the integral (\ref{eq:integral})
is restricted to an infinitesimal segment centered on the root $t_r$, since everywhere else 
the Dirac delta function is zero.

Inside this infinitesimal segment the function $g(t)$ is equal to its first order Taylor series approximation,
which is
\begin{equation}
g(t) = g(t_r) + (t - t_r) \frac{dg}{dt}\Big|_{t_r} = (t - t_r) \frac{dg}{dt}\Big|_{t_r} .
\label{eq:taylor}
\end{equation}

Step II. A change of variables gets the constant factor $|g'(t_r)|$ out of the Dirac delta function, 
a step equivalent to using the identity
\begin{equation}
\delta\big( a (x - x_o) \big) = \frac{1}{|a|} \delta(x - x_o),
\end{equation}
which is also easily proved with the same change of variables.

A similar procedure can be performed in Fourier space. 
Start by replacing the Dirac delta function in (\ref{eq:integral}) with
\begin{equation}
\delta\big( g(t) \big) = \frac{1}{2 \pi} \int e^{i \omega g(t)} \ d\omega.
\label{eq:fourier_in}
\end{equation}

Next, as in Step I, we restrict the domain of integration over $t$ to an 
 infinitesimal segment centered on the root $t_r$, and then we replace $g(t)$ with
 its equivalent expression from (\ref{eq:taylor}). The integral (\ref{eq:integral}) becomes
\begin{equation}
\int_{t_r - \epsilon}^{t_r + \epsilon} dt \ \frac{1}{2 \pi} \int d\omega \ e^{i \omega (t - t_r) g'(t_r)} \ f(t),
\label{eq:integral2}
\end{equation}
where $\epsilon$ is an infinitesimal number.

Next, as in Step II, we perform a change of variables, with $\omega' = \omega\ g'(t_r)$.
The integral (\ref{eq:integral2}) becomes
\begin{equation}
\int_{t_r - \epsilon}^{t_r + \epsilon} dt \ \frac{1}{2 \pi} \int d\omega' 
\ \frac{1}{g'(t_r)} \ e^{i \omega' (t - t_r)} \ f(t).
\label{eq:integral3}
\end{equation}

We now get out of Fourier space, using
\begin{equation}
\delta(t - t_r) = \frac{1}{2 \pi} \int e^{i \omega' (t - t_r)} \ d\omega'.
\label{eq:fourier_out}
\end{equation}
The integral (\ref{eq:integral3}) becomes
\begin{equation}
\int_{t_r - \epsilon}^{t_r + \epsilon} dt \ \frac{1}{g'(t_r)} \ \delta(t - t_r) \ f(t).
\label{eq:integral4}
\end{equation}

Finally, because of the Dirac delta function, we could extend the domain of integration over $t$
to its original full range, thus proving the identity (\ref{eq:Dirac_delta_property}) in the simple case considered.

As a result of this mathematical analysis, we disagree with the assertion that
a direct use of the identity (\ref{eq:Dirac_delta_property})
represents \lq\lq a lost opportunity to bring out the physics\rq\rq\ \cite{Perosa2023}.

The authors of Ref. \cite{Perosa2023} have also derived the standard expression of the Doppler factor,
and this derivation also deserves further clarifications. 

In the following part of this Comment, unless explicitly mentioned otherwise, all equation numbers
refer to equations from Ref. \cite{Perosa2023}.

This is our additional list of more detailed observations:

$\bullet$ The name of David J. Griffiths is misspelled, the last letter is missing.

$\bullet$ Typo in the paragraph under Eq. (4), 
instead of $(t + r/c)$ we have $\delta(t + r/c)$.

$\bullet$ Typo in Eq. (5), instead of $e^{i \omega t}$ we have $e^{- i \omega t}$.

$\bullet$ The authors use ${\bf r}$ as the position vector of the field point.
This is the place where the L-W potentials are calculated, a fixed point $P$ in the reference frame
shown in Fig. 1. As such, ${\bf r}$ is constant. 

Accordingly, $\frac{d{\bf r}}{dt} = 0$, and not $\frac{d{\bf r}}{dt} = {\bf v}$, 
as seen in Eq. (9).

$\bullet$ In Eq. (3) the authors use ${\bf r}_0$ as the position vector of the moving charged particle.
As such, ${\bf r}_0$ changes with time.
We therefore expect to have an analogy 
with the Doppler effect characterizing a moving source and an observer at rest, and the authors even
mention the fact that a \lq\lq relevant consequence of the motion of the source is the {\it Doppler shift}
of the wave frequency\rq\rq.
This described situation is not in agreement with Eq. (6), 
since this equation applies to the case of a source at rest at the origin.

$\bullet$ If Eq. (6) is replaced with the correct equation
\begin{equation}
\phi({\bf r}, t) = {\bf k} \cdot \big({\bf r} - {\bf r}_0(t_r)\big) - \omega t,
\end{equation}
then 
\begin{equation}
\phi({\bf r}_0(t_r), t_r) = {\bf k} \cdot \big({\bf r}_0(t_r) - {\bf r}_0(t_r)\big) - \omega t_r = - \omega t_r,
\end{equation}
and the unnumbered equation between Eqs. (7) and (8) makes sense
\begin{equation}
\phi({\bf r}, t) - \phi({\bf r}_0(t_r), t_r) = {\bf k} \cdot \big({\bf r} - {\bf r}_0(t_r)\big) - \omega (t - t_r) = 0,
\end{equation}
provided that $k = \omega/c$ and that
\begin{equation}
\hat{\bf k} = \frac{{\bf r} - {\bf r}_0(t_r)}{|{\bf r} - {\bf r}_0(t_r)|},
\end{equation}
and not $\hat{\bf k} = \hat{\bf r}$, as seen in Eq. (10), since this expression 
applies to the case of a static source at the origin.
The correct formula for $\hat{\bf k}$ shows up in Eq. (15), where $\hat{\bf k}$ is renamed $\hat{\bf R}$.
This inconsistent notation is undeniable in Eq. (18), where we have both $\hat{\bf R}$ and $\hat{\bf k}$.

$\bullet$ How would a correct derivation of the Doppler factor look like?
The expression of the Doppler factor is $\frac{dt_r}{dt}$,
as derived by Kapoulitsas \cite{Kapoulitsas1981} based on the changes of phase at source and detector.
This important expression is mentioned in the Appendix, in Eq. (A.7).

From the expression (8) of the retarded time one can write $t$ as a function of $t_r$
\begin{equation}
t = t_r + \frac{|{\bf r} - {\bf r}_0(t_r)|}{c},
\end{equation}
and then take the derivative, in order to find the reciprocal of the Doppler factor
\begin{equation}
\frac{dt}{dt_r} = 1 + \frac{1}{c} \frac{d\ |{\bf r} - {\bf r}_0(t_r)|}{dt_r}
= 1 - \frac{{\bf v}(t_r) \cdot \big({\bf r} - {\bf r}_0(t_r)\big)}{c \ |{\bf r} - {\bf r}_0(t_r)|}
= \frac{1}{D({\bf v}, \hat{\bf k})},
\end{equation}
where ${\bf v}$ is the velocity of the source. From this equation we also obtain that
\begin{equation}
\frac{d\ |{\bf r} - {\bf r}_0(t_r)|}{dt_r} = c \left( \frac{dt}{dt_r} - 1 \right).
\end{equation}

The phase can be written as
\begin{equation}
\phi({\bf r}, t) = \frac{\omega}{c} |{\bf r} - {\bf r}_0(t_r)| - \omega t,
\end{equation}
and the Doppler shifted frequency can be found as suggested by the authors in Eq. (9), however
please notice that, for a moving source, 
the binomial $(1 - \hat{\bf k} \cdot {\bf v} / c)$ should divide, not multiply, $\omega$.
In other words, at the end of Eq. (9), $D({\bf v}, \hat{\bf k})$ should be a factor, not a divisor,
in agreement with the position that it has in the L-W electric potential (18).

Since $t_r$ is a function of $t$
\begin{equation}
- \omega'= \frac{d\phi}{dt} = \frac{\omega}{c} \frac{d\ |{\bf r} - {\bf r}_0(t_r)|}{dt} - \omega
= \frac{\omega}{c} \frac{d\ |{\bf r} - {\bf r}_0(t_r)|}{dt_r} \frac{dt_r}{dt} - \omega
= - \omega \frac{dt_r}{dt},
\end{equation}
and this completes the proof of the Doppler effect formula.

$\bullet$ The calculation of the L-W potentials is obscured by the fact that, in Eqs. (5) and (13), 
the authors use $k$ instead of $\omega/c$.
Even though $k = \omega/c$, the independent integration variable is $\omega$.
After the integration over $\omega$ in Eq. (13), we get the Dirac delta function seen in Eq. (12).
This Dirac delta function ensures the fact that $t'= t_r$. This condition does
not happen before the integration over $\omega$. 
Nonetheless, the authors assume that $t'= t_r$ when they go from the middle
to the bottom line in Eq. (14). 
\begin{equation}
\phi({\bf r}, t) = k |{\bf r} - {\bf r}_0(t_r)| - \omega t,
\end{equation}
\begin{equation}
\phi({\bf r}_0(t'), t') = k |{\bf r}_0(t') - {\bf r}_0(t_r)| - \omega t',
\end{equation}
\begin{equation}
\phi({\bf r}, t) - \phi({\bf r}_0(t'), t') = k |{\bf r} - {\bf r}_0(t_r)| - k |{\bf r}_0(t') - {\bf r}_0(t_r)| - \omega (t - t'),
\label{eq:deltaphi}
\end{equation}
which is equal to 
\begin{equation}
k |{\bf r} - {\bf r}_0(t')| - \omega (t - t'),
\end{equation} 
when $t' = t_r$, when both expressions above are equal to zero.

$\bullet$ One could also notice that
\begin{equation}
k |{\bf r} - {\bf r}_0(t')| - \omega (t - t') = \omega\ g(t'),
\label{eq:omegag}
\end{equation}
where the function $g(t')$ was defined in Eq. (\ref{eq:function}) of this Comment, and accordingly also define
\begin{equation}
\phi({\bf r}, t) - \phi({\bf r}_0(t'), t') = \omega\ h(t').
\label{eq:omegah}
\end{equation}
By performing the integration over $\omega$ before the integration over $t'$,
a Dirac delta function appears, and as a consequence
the range of integration over $t'$ can be restricted to an infinitesimal segment centered on $t_r$.
Both functions $g(t')$ and $h(t')$ are equal to zero when $t' = t_r$.
Is this enough to allow us to replace $g(t')$ with $h(t')$? The answer is no, because $g(t')$ and $h(t')$ are
arguments of a Dirac delta function.
Because of the composite Dirac delta function, what matters is not only the
value of the function inside the Dirac delta function, but also the value of its derivative.
We must make sure that the derivatives $g'(t_r)$ and $h'(t_r)$ are also equal.

$\bullet$ As already mentioned, 
a standalone derivation of Eq. (20), in the simple case when the function $g(t)$ has only one root $t_r$, 
with $g(t_r) = 0$ and $g'(t_r) \neq 0$, consists of two steps.

Step II is performed in Eq. (16), where the two Dirac delta functions 
$\delta\big( a (t' - t_r) \big)$ and $\delta(t'- t_r)$
have been replaced with their Fourier representations before the change of variables is made, 
with $a = 1/D({\bf v}, \hat{\bf R}) = g'(t_r)$.

Step I is not as clearly explained. The authors
mention that something must be expanded around $t_r$, but this is not the phase difference,
as stated, since the phase difference 
shows up in Eq. (14) without proper mathematical justification, 
the quantity to be expanded around $t_r$ is the previous exponent, 
whose imaginary part is given in Eq. (\ref{eq:omegag}) of this Comment.
It is now that the exponent becomes $i \omega (t - t_r) g'(t_r)$, 
as seen in Eq. (\ref{eq:integral2}) of this Comment.
Step I is thus performed when we go from the middle line of Eq. (14) to the top line of Eq. (16).

$\bullet$ The authors mention that only close to $t_r$ we get 
relevant contributions in the integration over $\omega$,
this is incorrect. The integration is over $t'$.

$\bullet$ The calculation of $g'(t_r)$ produces the reciprocal of the Doppler factor.
\begin{equation}
\frac{dg}{dt'}\Big|_{t_r} = 1 + \frac{1}{c} \frac{d \ |{\bf r} - {\bf r}_0(t')|}{dt'}\Big|_{t_r}
= 1 + \frac{1}{c} \frac{d \ |{\bf r} - {\bf r}_0(t_r)|}{dt_r}  = \frac{dt}{dt_r}
= \frac{1}{D({\bf v}, \hat{\bf R})}.
\end{equation}

$\bullet$ The calculation of $h'(t_r)$ produces a result that is different from $g'(t_r)$.
From Eqs. (\ref{eq:deltaphi}) and (\ref{eq:omegah}) of this Comment we can see that
\begin{equation}
h(t') = \frac{1}{c} |{\bf r} - {\bf r}_0(t_r)| - \frac{1}{c} |{\bf r}_0(t') - {\bf r}_0(t_r)| - t + t'.
\end{equation}
Thus
\begin{equation}
\frac{d\ h(t')}{dt'} = 1 - \frac{1}{c} \frac{d\ |{\bf r}_0(t') - {\bf r}_0(t_r)|}{dt'}.
\end{equation}
Since
\begin{equation}
|{\bf r}_0(t') - {\bf r}_0(t_r)| = \sqrt{[x_0(t') - x_0(t_r)]^2 + [y_0(t') - y_0(t_r)]^2 + [z_0(t') - z_0(t_r)]^2},
\end{equation}
a straightforward application of the rules of calculus gives
\begin{equation}
\frac{d\ |{\bf r}_0(t') - {\bf r}_0(t_r)|}{dt'} 
= \frac{{\bf v}(t') \cdot \big({\bf r}_0(t') - {\bf r}_0(t_r)\big)}{|{\bf r}_0(t') - {\bf r}_0(t_r)|}.
\end{equation}

Care must be taken now, because we cannot simply substitute $t' = t_r$ in the expression above. Instead,
we consider a limiting process, $t' \to t_r$, that leaves us 
with $|{\bf v}(t_r)|$ when $t' \downarrow t_r$,
and with $- |{\bf v}(t_r)|$ when $t' \uparrow t_r$.

Since $h'(t_r) \neq g'(t_r)$, 
the last line in Eq. (14) and the first part of Eq. (15) have been mistakenly included 
here, where they are not even needed. 

$\bullet$ Typo in the line above Eq. (21), instead of ${\bf r'}(t')$ we have ${\bf r}_0(t')$ inside the absolute value bars.

\end{document}